\documentclass[sigplan, nonacm]{acmart}

\usepackage{tikz}
\usepackage{cleveref}
\usepackage{amsmath}

\usepackage{filecontents}

\usepackage{algorithmic}
\usepackage{graphicx}
\usepackage{textcomp}
\usepackage{xcolor}

\definecolor{indigo}{RGB}{51,0,102}
\usepackage{tcolorbox}

\usepackage{hyperref}

\usepackage[framemethod=TikZ]{mdframed}

\usepackage{xspace}
\usepackage{scalerel}
\usepackage{ctable}

\usepackage{pifont}

\newcommand{\neuroscalar}{NeuroScalar{}}

\usepackage{url}
\usepackage{mathtools}

\usepackage{tabularray}

\usepackage{multirow}
\usepackage{array}
\usepackage[normalem]{ulem}

\AtBeginDocument{%
  }
\copyrightyear{2025}

\usepackage{graphicx} 

\title{NeuroScalar: A Deep Learning Framework for Fast, Accurate, and In-the-Wild Cycle-Level Performance Prediction}
\author{Shayne Wadle, Yanxin Zhang, Vikas Singh, Karthikeyan Sankaralingam}
\affiliation{%
  \institution{University of Wisconsin - Madison}\city{Madison}\country{USA}}

\begin{abstract}
The evaluation of new microprocessor designs is constrained by slow, cycle-accurate simulators that rely on unrepresentative benchmark traces. This paper introduces a novel deep learning framework for high-fidelity, ``in-the-wild'' simulation on production hardware. Our core contribution is a DL model trained on \textbf{microarchitecture-independent features} to predict cycle-level performance for hypothetical processor designs. This unique approach allows the model to be deployed on existing silicon to evaluate future hardware. We propose a complete system featuring a lightweight hardware trace collector and a principled sampling strategy to minimize user impact. This system achieves a simulation speed of \textbf{5 MIPS} on a commodity GPU, imposing a mere \textbf{0.1\% performance overhead}. Furthermore, our co-designed \textbf{Neutrino} on-chip accelerator improves performance by \textbf{85x} over the GPU. We demonstrate that this framework enables accurate performance analysis and large-scale hardware A/B testing on a massive scale using real-world applications.

\end{abstract}

\begin{document}

\maketitle
\pagestyle{plain}

\section{Introduction}
The relentless pace of innovation in microprocessor design is fundamentally constrained by the methodologies used to evaluate new architectural ideas. The gold standard, cycle-level simulation~\cite{akram2019survey,austin2002simplescalar,binkert2011gem5,carlson2011sniper,elrabaa2017very,patel2011marss,sanchez2013zsim}, provides high-fidelity performance data but suffers from prohibitive performance overheads. For instance, a detailed out-of-order CPU model in a widely-used simulator like gem5 may only achieve a simulation rate of 0.1 million instructions per second (MIPS), several orders of magnitude slower than native hardware execution. This bottleneck is exacerbated by the ever-increasing complexity of modern processors. This, in turn, necessitates a correspondingly larger exploration of the design space. Furthermore, traditional evaluation relies heavily on standardized benchmark suites or curated workload traces. While useful, these benchmarks often fail to capture the dynamic and  unpredictable nature of ``in-the-wild'' applications that run on user devices that result from complex interactions. 
An ideal evaluation methodology would therefore possess three key properties: \textbf{speed} approaching native execution, \textbf{fidelity} matching that of a cycle-accurate simulator, and \textbf{transparency} that allows for evaluation on real user workloads without interference or noticeable slowdown. 

No existing technique successfully achieves this trifecta. While prior work has explored machine learning for performance prediction, notable approaches fall short of enabling true in-the-wild evaluation. For example, SimNet~\cite{10.1145/3530891} requires detailed microarchitectural traces as input, which are unavailable on production hardware for a hypothetical design. Other approaches, such as TAO~\cite{10.1145/3656012}, pursue the ambitious goal of learning a general model of processor behavior but focus on ``high-level'' hardware trends. This is of limited value to a chip designer as these high-level
trends are often already known. The true secret sauce of modern microarchitecture lies in extreme feature engineering---subtle choices like prefetcher policies, load-store queue (LSQ) sizing, or the number of load ports. TAO is ill-equipped to capture the impact of such low-level details; its authors concede that accuracy suffers on branch-predictor-dependent applications. Furthermore, its massive model size precludes any practical form of the sampling-based, in-the-wild inference necessary for transparent deployment. 
This leaves a critical gap: a methodology to rapidly evaluate how specific,
low-level microarchitectural trade-offs perform on real-world workloads. 

This paper addresses this multifaceted challenge by introducing a novel paradigm for ultra-fast, in-the-wild simulation of new microarchitectural concepts on existing production silicon. Our approach is centered on a deep learning (DL) model trained \textit{a priori} to perform cycle-level performance prediction. The key insight is to train this model using only microarchitecture-independent features---that is, features derived solely from the instruction stream and its data dependencies. During deployment, this pre-trained model runs on production chips in the hands of users, ingesting instruction traces from the host processor and predicting the execution latencies of a new, hypothetical microarchitecture. As even the DL model inference is too slow to process constantly produced data at native speed, we employ a sampling-based methodology. A lightweight hardware monitor captures a sample epoch of instructions (e.g., 100,000 instructions), which is then processed by the DL model. Our model achieves an inference rate of \textasciitilde4 MIPS on desktop class GPUs like RTX4090. By tuning the sampling frequency---for instance, sampling one epoch every 75 billion instructions (a 100,000 instruction epoch every \textasciitilde25 seconds)---we can scavenge idle resources from an accompanying GPU, introducing a negligible GPU slowdown of just 0.1\%. When using our co-designed 28mW Neutrino in-silicon accelerator, this sampling rate dramatically increases to an epoch every 4.9 seconds. With an 8-tile design, further reducing to an epoch every 0.6 seconds, while consuming 230mW.

This framework, we call \neuroscalar{} enables powerful new modalities for architectural evaluation. It can be used to gain deep insights into how diverse, real-world applications perform on a proposed design: an \textit{Enterprise Forecaster} like a hyperscalar who wants to understand future chip (their own or from a vendor like AMD, Intel, NVIDIA) on their proprietary workloads. More significantly, for chip designers (who are not hyperscalars - Intel, AMD, NVIDIA) it facilitates large-scale design space exploration by allowing multiple candidate designs (represented by different trained models) to be evaluated concurrently on live workloads (through Ecosystem Partners), creating the hardware equivalent of A/B testing. The primary contributions of this work are: 
\begin{enumerate} 
\item The design and implementation of a fast and accurate DL model for cycle-level performance prediction that relies on microarchitecture-independent features.
\item The design of a complete system for in-field evaluation, including a simple hardware module for trace collection and an ultra-lightweight, 28mW accelerator for the DL model. This accelerator enables continuous sampling and achieves an 85$\times$ energy reduction and a 391$\times$ area reduction compared to GPU-based inference, while allowing a 5$\times$ higher sampling rate.
\item A thorough evaluation of the model's predictive accuracy against a cycle-level simulator.
\item A demonstration of the framework's effectiveness for design space exploration, where it achieves 95\% accuracy in A/B testing across eight pairwise comparisons of five distinct processor designs. 
\end{enumerate} 

This paper is organized as follows. Section 2 provides an overview of our approach. Section 3 details the DL model architecture. Section 4 describes the end-to-end system design, including the hardware components. Section 5 outlines our experimental methodology, and Section 6 presents our results. We discuss related work in Section 7 and conclude.

\section{System Overview}

This section provides a high-level overview of the NeuroScalar system. 
We frame the core challenges in processor simulation that motivate our work 
and present our key technical insights. We describe the end-to-end flow, defining the roles of its users and the practical use cases it enables.

\subsection{Challenges and Key Insights}

The central challenge in modern microprocessor design is the three-way trade-off between simulation \textbf{fidelity}, \textbf{speed}, and \textbf{workload relevance}. Traditional cycle-level simulators provide high fidelity but are excruciatingly slow (often by a factor of 1,000,000x), rendering large-scale design space exploration impractical. This speed limitation also forces designers to rely on standardized benchmark traces, which fail to capture the rich, data-dependent behaviors of the ``in-the-wild'' applications that users run on a daily basis. \textit{Our work directly confronts this challenge: how can we evaluate novel microarchitectural ideas with cycle-level accuracy, at near-native speed, on real user workloads, and with complete transparency to the end-user?}

Our solution, NeuroScalar, is built upon a foundational insight: we can decouple the slow, offline process of detailed simulation from a fast, online prediction phase by using a deep learning (DL) model trained on \textbf{microarchitecture-independent features}. This allows a model, trained once by a chip designer in a laboratory setting, to be deployed on existing production silicon to predict the performance of a 
different, hypothetical processor. This breaks the dependency on slow, monolithic simulation for every design change.

A second key insight enables practical, ``in-the-wild'' deployment. To be viable, the prediction process must be non-intrusive. While prior work like Simnet and Tao demonstrated the potential of DL for simulation, their models were often too heavyweight for transparent deployment. NeuroScalar is explicitly co-designed to be \textbf{ultra-lightweight}. We combine a compact, 1-million-parameter model with an \textbf{sampling methodology}. Any high-fidelity simulation requires significantly more computation per instruction than native execution. For instance, even an idealized linear model on a state-of-the-art A100 GPU running at 100\% compute utilization, consuming 300W, cannot process instruction features as fast as a single CPU core produces them. Therefore, a principled sampling methodology is not a workaround, but a fundamental requirement for any transparent, in-the-wild simulation. Our work is architected around this reality. 
For scenarios without a powerful GPU, we also introduce a co-designed \textbf{hardware accelerator} that delivers high-speed inference within a tiny power budget of \textbf{28mW} at 7nm based on a full RTL and P\&R flow. Our cross-cutting insight is: \textit{processor simulation is fundamentally a time-series behavior, which LSTMs are intuitively well suited, for high accuracy they benefit from a context that is substantially larger than the ROB size, and a hardware accelerator can be co-designed to match the embedding dimension of the LSTM to achieve ultra-high efficiency. Specifically feature engineering to preserve address information, specializing layers for long and short sequences, log-transform to handle the wide range of output retirement cycles, a sequence length upto $3\times$ larger than an ROB demonstrates small models can accurately learn using microarchitecture-independent features only.}


\subsection{End-to-End Workflow and Use Cases}

The NeuroScalar system operates in a two-phase workflow that involves two key actors: the \textbf{chip designer} and the \textbf{end-user}. During the offline Training Phase, the \textbf{chip designer} performs a one-time, offline training process. Using a conventional cycle-accurate simulator, they generate training data to produce a suite of NeuroScalar models. 
Each lightweight model is a proxy for a specific microarchitectural idea (e.g., a new cache prefetch algorithm or branch predictor), effectively encapsulating a complex design into a portable neural network. During the \textit{online Inference Phase}, the trained models are distributed to end-users. A lightweight tracing mechanism on the user's machine---either a minor hardware modification or enhancing existing debug interface (e.g., Intel PT)---captures the required architectural features from running applications. These traces are then fed into the NeuroScalar model to predict cycle counts for the hypothetical hardware, based on a sampling strategy. This workflow enables powerful new use cases tailored to different users.

\begin{figure}[tbp]
    \centering
    \includegraphics[width=\columnwidth]{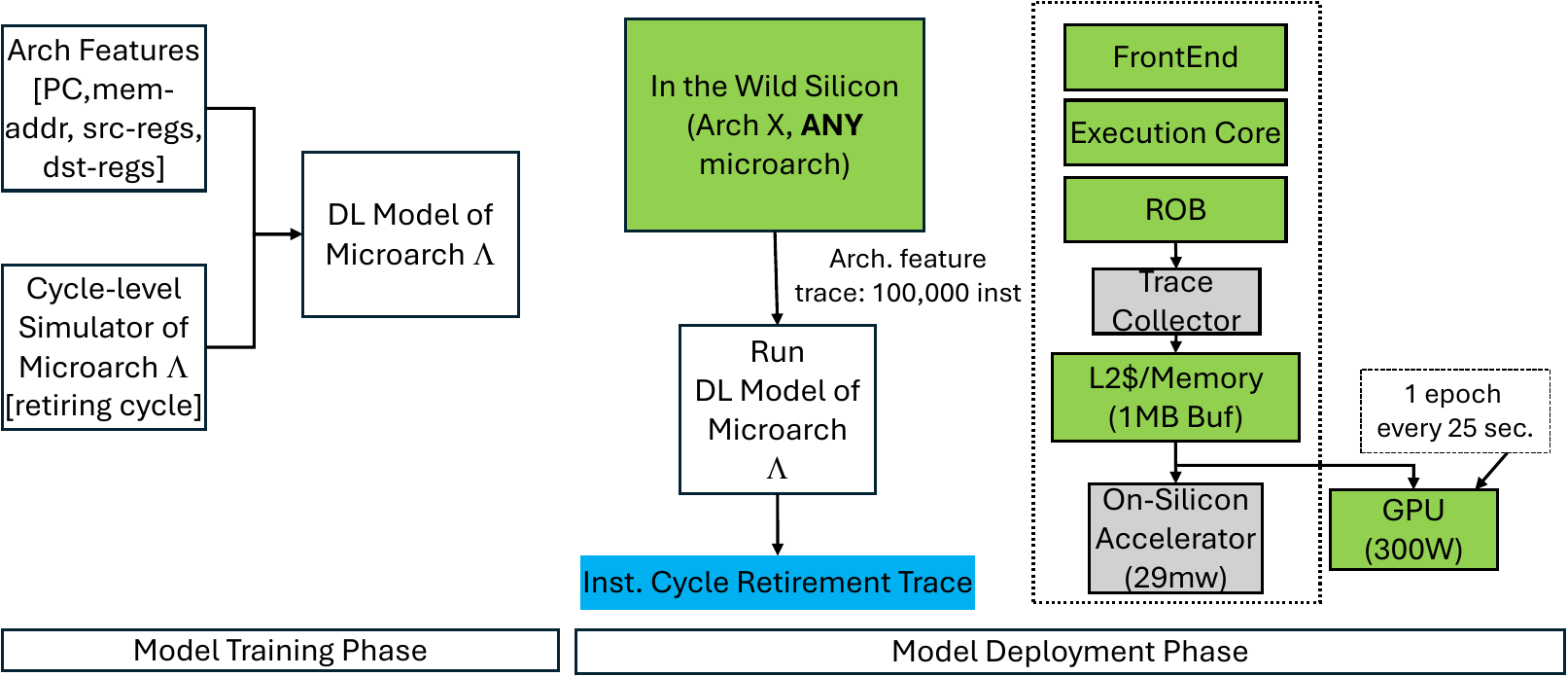}
    \caption{The NeuroScalar end-to-end workflow, showing the offline training phase performed by the chip designer and the online inference on the end-user's system.}
    \label{fig:workflow}
\end{figure}

\begin{itemize}
\item \textbf{For the Enterprise Forecaster:} A customer, like a hyperscalar vendor, with proprietary workloads can use a NeuroScalar model to get a concrete performance forecast for a future processor on their own applications, all without sharing their sensitive code or data with the chip designer.
\item \textbf{For the Chip Designer:} The designer can now perform massive, in-the-wild A/B testing. By collecting anonymized performance reports from opt-in \textbf{Ecosystem Partners}, they can gain unprecedented insight into how their design ideas perform across a diverse, real-world software ecosystem. 
\end{itemize}

\section{DL Model Theory}

This section details the methodology for our cycle-level prediction framework. The data pipeline encodes each instruction with 13 microarchitecture-independent features. There are six key properties gathered from an instruction: PC, opclass, mem\_addr, src\_reg1, src\_reg2, and dst\_reg. The 64-bit PC and mem\_addr are represented as 3 features of 2x22bits + 20bits. Registers are represented as a pair (class and number). This accounts for the 13 features. 
They characterize ground-truth (GT) cycle distributions across benchmarks, and we mitigate severe target skew via clipping, a logarithmic transform, and an auxiliary two-way classifier. We then assess four neural architectures with respect to prediction accuracy, computational efficiency, and memory footprint: LSTM, CNN, SSM and Transformer based designs. We select an LSTM backbone based on its lightweight architecture and performance. An LSTM is a recurrent neural network architecture that utilizes a dedicated cell state and 
gating structures (input, forget, and output) to selectively manage the flow of sequential information, enabling it to model long-range dependencies~\cite{lstmdemo} -- a good fit here. Finally, we describe the predictor’s architecture, including a 
study to identify the optimal number of LSTM layers, and a unified design that jointly trains the regression and classification heads. 

\subsection{Data Representation and Preprocessing}

\subsubsection{Feature Selection}
Designing an effective cycle-level prediction framework requires features that not only reflect instruction-level semantics but also capture execution-relevant microarchitectural patterns. A central challenge lies in constructing a representation that is sufficiently expressive to encode spatial locality, operational semantics, and data dependencies, while remaining independent of specific microarchitectural configurations. Such a representation must generalize across workloads and hardware designs, avoid overfitting to architecture-specific idiosyncrasies, and still preserve information critical for latency prediction.

To address these requirements, we adopt a set of six carefully selected microarchitecture-independent properties, which together provide a holistic view of each instruction’s execution context. These key properties are represented in 13 features. These features fall into two categories: (1) continuous-valued encodings of instruction and memory addresses to preserve spatial locality, and (2) categorical descriptors such as operation type and register usage to convey semantic and dependency information. This combination enables the model to capture both fine-grained numerical patterns and discrete structural relationships, thereby mitigating the limitations of using either type alone.

\subsubsection{Feature Processing and Usage}
We next consider how each feature is encoded, transformed, and integrated into the model. Registers are encoded as a pair: their class and number. 

\paragraph{Address decomposition.}
Given a 64-bit address \(a\), we preserve locality and expose large jumps by a 3-way split: 

\[
\phi_{\mathrm{addr}}(a)=\Big(\underbrace{\lfloor a/2^{42}\rfloor}_{\text{upper 22b}},\;
\underbrace{\lfloor(a\bmod 2^{42})/2^{20}\rfloor}_{\text{middle 22b}},\;
\underbrace{a\bmod 2^{20}}_{\text{lower 20b}}\Big)\in\mathbb{R}^{3}.
\]

These are encoded as fp32 in order to preserve the entire bit-sequence; fp32 has 23 bits of precision in the mantissa. 

\paragraph{Embeddings and projections.}
Categorical features are embedded; continuous features are linearly projected into the same space:
\[
\begin{aligned}
& e^{\mathrm{op}}_t=\mathbf E_{\mathrm{op}}[o_t],\\
& e^{\mathrm{reg}}_{t,j}=\big[\mathbf E_{\mathrm{cls}}[c_{t,j}];\ \mathbf E_{\mathrm{idx}}[n_{t,j}]\big],\quad j\in\{\mathrm{dst},\mathrm{src1},\mathrm{src2}\},\\
& e^{\mathrm{pc}}_t=\mathbf W_{\mathrm{pc}}\phi_{\mathrm{addr}}(a^{\mathrm{pc}}_t),\quad
e^{\mathrm{mem}}_t=\mathbf W_{\mathrm{mem}}\phi_{\mathrm{addr}}(a^{\mathrm{mem}}_t).
\end{aligned}
\]
The instruction vector is the concatenation
\[
x_t=\big[e^{\mathrm{op}}_t;\ e^{\mathrm{reg}}_{t,\mathrm{dst}};\ e^{\mathrm{reg}}_{t,\mathrm{src1}};\ e^{\mathrm{reg}}_{t,\mathrm{src2}};\ e^{\mathrm{pc}}_t;\ e^{\mathrm{mem}}_t\big]\in\mathbb{R}^{13}.
\]

\paragraph{Sliding window formation.}
Training data is generated via a sliding window of arbitrary length \(N\). Given a target segment length \(R\le N\), we center the target and use the remaining positions as context. Let \(s=\lfloor (N-R)/2\rfloor\). Then
\begin{equation}\label{eq:win-split}
x_{1:N}=\big[x_{1:s}\ \|\ x_{s+1:s+R}\ \|\ x_{s+R+1:N}\big],\quad
\mathbf y=\mathbf y_{s+1:s+R}\in\mathbb{R}^{R},
\end{equation}
so the left context has length \(s\) and the right context has length \(N-R-s\). The model predicts cycles for the middle segment using both the preceding and succeeding context. All features are normalized before training, and the target cycles are preprocessed as in Section~\ref{sec:gt-process}. This is one of our fundamental insights, creating a very long context, we found $3\times$ of ROB size, provides accuracy while using only \textbf{microarchitecture-independent} features.

\begin{figure}[t]
  \centering
  \includegraphics[width=\linewidth]{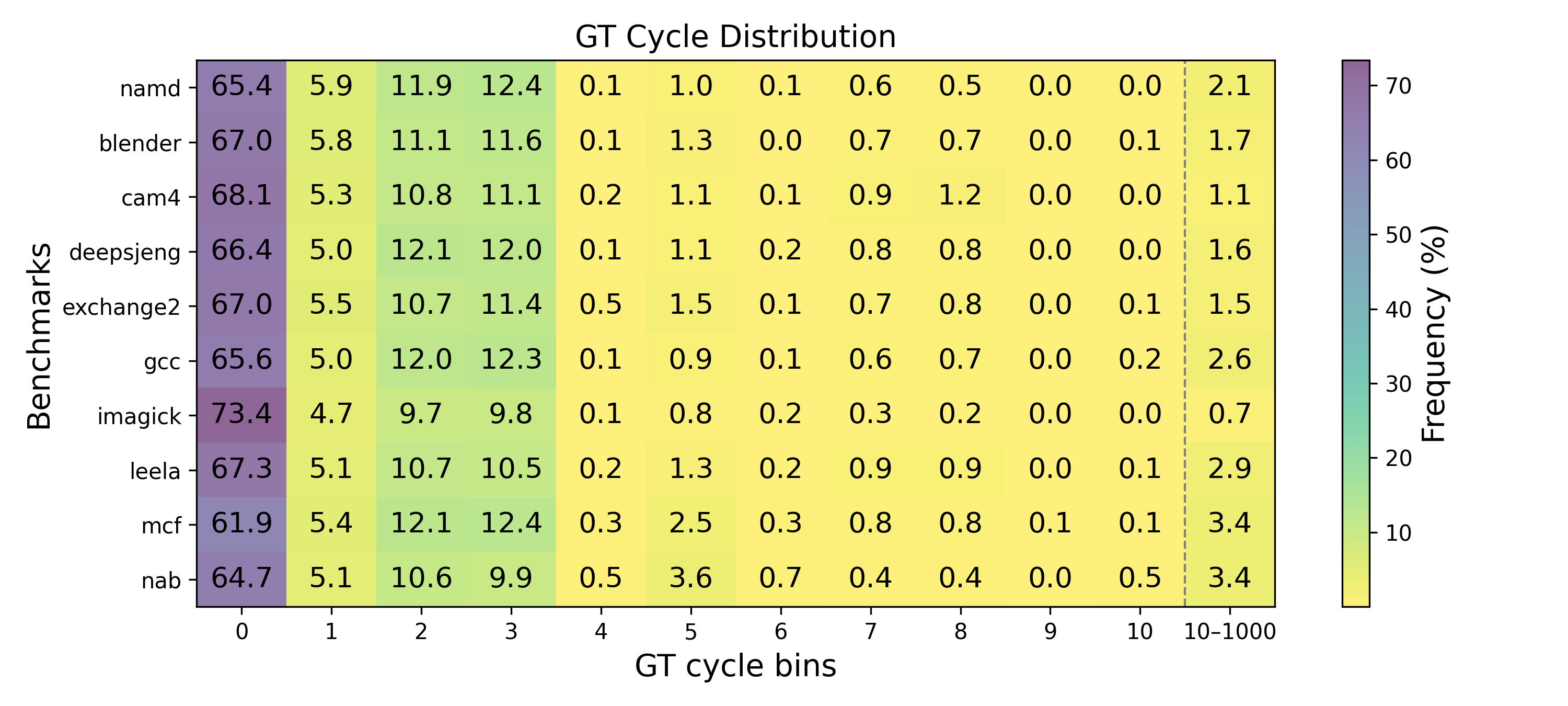}
  \caption{GT cycle distributions as percentage of total number of instructions shown per benchmark - the rows.}
  \label{fig:gt-dist-heatmap}
\end{figure}

\subsubsection{Ground Truth Distribution}
We examine the ground-truth (GT) cycle distribution for ten representative benchmarks: namd, blender, cam4, deepsjeng, exchange2, gcc, imagick, leela, mcf, and nab. The distribution for a given benchmark is calculated as the portion of the instructions of said benchmark that have a specified cycle latency. The distributions are visualized as a heatmap (Fig.~\ref{fig:gt-dist-heatmap}) where each row is a benchmark and columns represent integer cycle bins from 0 to 10, plus a long-tail bin covering 10 to 1000 cycles.

Across all benchmarks, the majority of instructions have very small cycle counts (mode at $0$ cycles and high density for 1,2, and 3 cycles), with more than $60\%$–$73\%$ of events occurring at zero cycles and another $20\%$–$25\%$ within the first three cycles. Beyond this dense low-latency region, the distributions exhibit a long but thin tail: bins beyond $10$ cycles account for only $0.7\%$–$3.4\%$ of number of instructions across benchmarks. Events beyond 1000 cycles are exceedingly rare. This extreme concentration in the low-latency region creates a challenging imbalance: regressors risk being dominated by the abundant short-latency samples while underrepresenting the patterns that characterize rare, high-latency cases.

\subsubsection{Ground Truth Processing}
\label{sec:gt-process}
To mitigate the effects of this heavy skew and improve regression performance, we apply a logarithmic transformation to the GT values. Concretely, we cap extremes and then compress the range:
\begin{equation}
\tilde y_i=\min(y_i,1000),\qquad z_i=\log\!\big(1+\tilde y_i\big).
\label{eq:log-transform}
\end{equation}

This compresses the dynamic range of the target variable, smooths the distribution, and reduces the disproportionate influence of rare high-latency samples, enabling the model to learn from a more balanced signal.

Furthermore, we modify the regression task with an auxiliary two-way classification task, in which each sample is labeled according to whether its GT cycle count falls below or above a 
threshold. With a tunable threshold $\tau$ (set to $\tau{=}10$ in our experiments), we define
\begin{equation}
c_i=\mathbf{1}\!\{\,y_i>\tau\,\}\in\{0,1\}.
\label{eq:regime-label}
\end{equation}
This choice aligns with the natural boundary between the dense low-latency region and the sparse long tail, and the framework allows $\tau$ to be adjusted for different application needs. Both the classification and regression tasks are trained jointly, sharing a common feature representation; this joint optimization enables the classification branch to dynamically inform the regression head about the likely latency regime during inference, avoiding the error-propagation risks of a sequential approach and supporting end-to-end learning that is both more accurate and more flexible in practice.

In summary, our 
main steps are: 
\textit{Address decomposition, large context, log-transform, and specialized long/short region predictors which are novel yet simple techniques that enable a small LSTM to effectively learn microarchitecture performance, without requiring a detailed microarchitecture trace.}

\subsection{Choice of Models}
We formulate cycle-level prediction as a supervised regression problem with a clear input–output association task. Given a window $x_{1:N}\in\mathbb{R}^{N\times 13}$ (Section~3.1), the model estimates the GT cycles for the centered $R$-length target segment (Eq.~\ref{eq:win-split}). In other words, we learn the mapping
\[
f_{\theta}:\ \mathbb{R}^{N\times 13}\rightarrow \mathbb{R}^{R},\qquad
\widehat{\mathbf y}=f_{\theta}(x_{1:N}),
\label{eq:seq2seq}
\]
augmented with an auxiliary two-way classifier that shares the backbone to indicate the low-/high-latency regime (Section~3.1.3).

To instantiate $f_{\theta}$, we evaluated four backbone families - LSTM, Transformer, CNN, and SSM - covering recurrent, attention-based, convolutional, and state-space designs. Transformers proved too heavy for our lightweight, real-time setting, while CNNs and SSMs underperformed. LSTM offered the best accuracy–efficiency trade-off and was therefore adopted as our backbone. Further analysis is in Section~3.3.

\begin{table*}[t]
  \centering
  \caption{Layer-depth study for the BiLSTM backbone. Speed is in million instructions per second (M instr/s).}
  \label{tab:lstm-depth}
  \setlength{\tabcolsep}{5pt}
  \begin{tabular}{lccccccccc}
    \toprule
    Layers $L$ & MAE$\downarrow$ & RMSE$\downarrow$ & RAE$\downarrow$ & Acc(round)$\uparrow$ & $\pm$1$\uparrow$ & $\pm$2$\uparrow$ & rel$\le$5\%$\uparrow$ & Speed (M)$\uparrow$ & val\_loss$\downarrow$ \\
    \midrule
    1 & 0.3623 & 4.9636 & 0.1387 & 0.7948 & 0.9512 & \textbf{0.9856} & 0.7998 & \textbf{4.77} & 0.0290 \\
    \textbf{2} & \textbf{0.3509} & \textbf{4.9386} & \textbf{0.1340} & \textbf{0.7968} & \textbf{0.9533} & 0.9845 & \textbf{0.8013} & 4.00 & \textbf{0.0283} \\
    3 & 0.3579 & 4.9984 & 0.1357 & 0.7895 & 0.9520 & 0.9840 & 0.7938 & 3.12 & 0.0287 \\
    \bottomrule
  \end{tabular}
\end{table*}

\subsection{Model Structure}
For our regression task -- a window of $N$ instruction-level feature vectors serve as the input and the GT cycles for the central $R$ positions are the output. We now ask: what computational scaffold most reliably turns contextual instruction streams into accurate cycle estimates? Our design proceeds in three steps: (i) construct a backbone that faithfully encodes sequential context seen before and after the target region, (ii) attach task heads that express both continuous (regression) and discrete (regime) structure of latency, and (iii) validate capacity via a depth study before finalizing the configuration for deployment.

\subsubsection{Backbone}
We adopt a stacked bidirectional LSTM (BiLSTM) as the sequence encoder. Since the prediction target is the centered segment within a sliding window, both past and future context are available during training and inference; thus, strict causality is not required in this setting. The input to the backbone is the length-$N$ sequence of instruction embeddings from Section~3.1, linearly projected to a hidden size $H$:
\[
x_{1:N}\in\mathbb{R}^{N\times 13},\qquad
\tilde x_t = \mathbf W_{\mathrm{in}} x_t + \mathbf b_{\mathrm{in}}\in\mathbb{R}^{H}.
\]
The BiLSTM (implemented with \texttt{batch\_first}) yields contextualized states by concatenating the forward and backward hidden vectors:
\[
\mathbf h_t=\big[\overrightarrow{\mathrm{LSTM}}(\tilde x_t);\ \overleftarrow{\mathrm{LSTM}}(\tilde x_t)\big]\in\mathbb{R}^{2H}.
\]
Stacking across time gives
\[
\mathbf H=[\mathbf h_1,\ldots,\mathbf h_N]^\top\in\mathbb{R}^{N\times 2H}.
\]
Then slice the centered target region as the shared representation for prediction, following the window split in Eq.~(\ref{eq:win-split}):
\begin{equation}
\mathbf H_{\mathrm{mid}}=\mathbf H_{\,s+1:s+R}\in\mathbb{R}^{R\times 2H}.
\label{eq:mid-slice}
\end{equation}
This matches our data formulation (Section~3.1.2), which leverages both left and right context to forecast the middle region.

\subsubsection{Heads and unified design}
On top of $\mathbf{H}_{\mathrm{mid}}$ (Eq.~\ref{eq:mid-slice}), we attach (i) a regression head that maps each hidden state to a scalar cycle prediction, and (ii) a two-way classifier that predicts whether the instruction lies in the short- or long-latency regime (threshold at 10 cycles; Section~3.1.3). Concretely, the classifier applies a two-layer MLP with ReLU to produce logits $\mathbf{p}\in\mathbb{R}^{R\times 2}$. For regression, we employ two lightweight heads, $g_{\text{short}}$ and $g_{\text{long}}$, each a single-layer MLP from $\mathbb{R}^{2H}\!\to\!\mathbb{R}$, and select per-instruction outputs by a mask induced from the predicted class. This decouples local response surfaces for different latency regimes while preserving a single shared encoder.

\subsubsection{Depth study (capacity vs.\ efficiency)}
Under identical training and evaluation settings, we vary $L\!\in\!\{1,2,3\}$. As shown in Table~\ref{tab:lstm-depth}, $L{=}2$ offers the best accuracy–efficiency trade-off with competitive throughput (4.00M instr/s) and the lowest validation loss (0.0283). By comparison, $L{=}1$ is faster (4.77M instr/s) but slightly less accurate, while $L{=}3$ yields no accuracy gains and incurs higher cost (3.12M instr/s). We therefore adopt $L{=}2$ as the default.

\subsubsection{Putting it together}
\begin{figure}[h]
    \centering
    \includegraphics[width=0.75\linewidth]{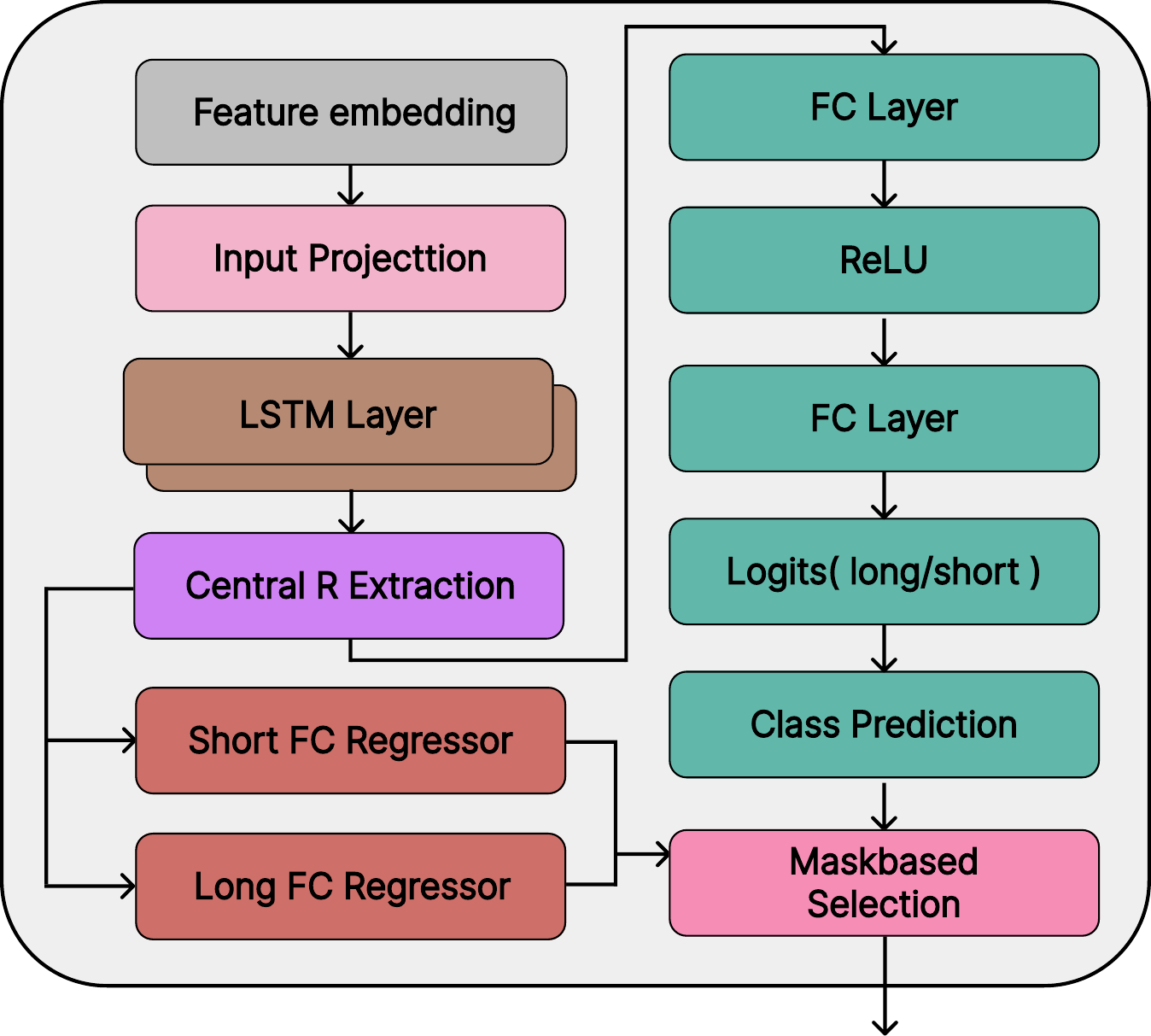}
    \caption{Overall architecture of the proposed LSTM-based cycle predictor.}
    \label{fig:model_structure}
\end{figure}

Figure~\ref{fig:model_structure} shows the final architecture. An input projection aligns heterogeneous features into a common $H$-dimensional space; an $L{=}2$ stacked BiLSTM encodes the window bidirectionally; the central $R$ embeddings feed a unified head block with a two-way classifier and regime-conditioned regressors. Unless stated otherwise, we set $H{=}128$.

\subsubsection{Training and joint objective}
Two practical challenges guide our training design: (a) the regression target is heavy-tailed even after clipping, and (b) the classifier and regressors must share signal without cascading errors. Thus, we adopt \emph{joint training} with a single loss that uses the log-transformed target $z_i$ (Eq.~\ref{eq:log-transform}) and the regime labels $c_i$ (Eq.~\ref{eq:regime-label}):
\[
\mathcal{L}
= \underbrace{\tfrac{1}{R}\sum_{i=1}^{R}\mathrm{SmoothL1}\big(\widehat{y}_i,\ \log(1+y_i)\big)}_{\text{regression on log-transformed cycles}}
\;+\;
\lambda\ \underbrace{\tfrac{1}{R}\sum_{i=1}^{R}\mathrm{CE}\big(\widehat{\mathbf{p}}_i,\ c_i\big)}_{\text{two-way classification}},
\]
The shared encoder lets the classifier shape the representation toward regime-discriminative features, while the regime-conditioned heads prevent a single regressor from being biased toward the dominant short-latency region. We train end-to-end with dropout on recurrent layers and early stopping on validation MAE; unless noted, the joint setting is used for all experimental results.

\subsection{Accuracy and Downstream Tasks}
Our design choices aim to improve predictive performance while remaining mindful of deployment constraints. Similar to vision models that must operate efficiently on edge devices, high ``raw accuracy'' is desirable but not the sole criterion of success. What ultimately matters is whether the backbone delivers strong performance in the context of downstream tasks. A model that achieves slightly higher raw accuracy at the expense of significantly greater computational cost may provide only marginal benefits for the end application.  

In this sense, our backbone plays a role 
akin 
to that of foundational encoders in vision and video processing. The raw accuracy of such models should be interpreted in light of their downstream utility. As 
our results show, 
although the backbone achieves 70–85\% raw accuracy, it enables downstream tasks—such as selecting between processor configurations across diverse benchmarks—to reach over 95\% accuracy. This demonstrates that the backbone strikes an effective balance between efficiency and task-level performance.

\section{System Design}

To realize the vision of in-the-wild microarchitectural simulation, a practical and robust system design is paramount. This section details the end-to-end architecture of \neuroscalar{}, focusing on the two central challenges: the efficient \textbf{collection of feature traces} and the \textbf{high-speed execution of model inference}. We address these challenges by presenting a complete data and execution pipeline. First, we describe the mechanisms for gathering the necessary microarchitecture-independent features, both from a cycle-level simulator during the offline training phase (Section~\ref{sec:cycle-gather}) and from live production hardware during online inference (Section~\ref{sec:live-gather}). Next, we detail two distinct deployment targets designed to run the \neuroscalar{} model with minimal overhead. We present a software-based inference engine optimized for commodity GPUs, designed to scavenge resources transparently. We then describe a co-designed, power-efficient on-chip accelerator for environments where a GPU is not available or where dedicated hardware is preferred. 

\subsection{Collecting Traces for Inference}
\label{sec:live-gather}
The first critical component of the \neuroscalar{} system is an efficient and low-overhead mechanism for collecting instruction traces from a processor running in a production environment. This section defines the composition of these traces, details our proposed hardware solution for capturing them, and addresses the key system-level challenges of integration and security.

\subsubsection{Trace and Epoch Definitions}
For the \neuroscalar{} model to predict performance, it requires a stream of microar-chitecture-independent features for each instruction. Our trace format is composed of six fundamental and readily available signals: the \textbf{Program Counter (PC)}, the \textbf{full memory address} for load/store operations, the \textbf{instruction opcode class}, and the register identifiers for \textbf{source register 1}, \textbf{source register 2}, and the \textbf{destination register}.

To manage data flow and enable batched processing, we group instructions into fixed-size units called \textit{epochs}. An epoch is defined as a contiguous block of \textbf{100,000 instructions}. Traces are collected for an entire epoch and stored in a buffer; then it is consumed by the inference engine, which allows for significant weight reuse and improved computational efficiency. The epoch size is a tunable parameter largely determined by feature dedicated memory capacity. 

While an epoch has no specific semantic meaning on its own, it is useful to understand the diversity of the code being executed. To this end, we compute a signature for each captured epoch by hashing its sequence of PCs. This signature allows an \textbf{Enterprise Forecaster} to analyze workload coverage, observe how different application inputs affect execution paths, and ensure a representative sample is collected over an application's runtime.

\subsubsection{Trace Collection Hardware and System Integration}
While existing technologies like binary instrumentation or hardware tracing mechanisms (e.g., Intel Processor Trace) can provide this data, they typically incur prohibitive performance overheads, making them unsuitable for a transparent, in-the-wild deployment.

Therefore, we propose a simple and minimally invasive hardware \textbf{trace collector} (See Figure~\ref{fig:accelerator-overview}). Our design consists of a small FIFO buffer attached to the processor's Re-Order Buffer (ROB). As instructions retire, the six required feature signals---which are readily available at this stage in a modern out-of-order processor---are written into the FIFO. To ensure proper operation in a multitasking environments, the trace collector is managed by the operating system. Its activation is tied to a process ID, ensuring that tracing is automatically paused during context switches and only resumes when the target application is scheduled, thus preventing contamination of the trace from other processes or the kernel.

The FIFO's output is mapped to system memory, allowing it to be drained through the conventional memory hierarchy. A FIFO with 512 entries, with each entry storing features for 5 instructions, results in a total on-chip trace buffer of just \textbf{12KB}. The memory storage needed for a 100,000 instruction-long epoch is 2.5MB. We summarize below a complete Power-Performance-Area (PPA) analysis:
\begin{itemize}
    \item \textbf{Area:} The area overhead is a negligible 12KB for the FIFO. We explicitly avoid a large, multi-megabyte dedicated SRAM, which would be prohibitively expensive.
    \item \textbf{Performance:} The processor front-end is stalled only if the buffer becomes full. This is a rare event, as modern L2 cache subsystems are well-equipped to absorb such write streams.
    \item \textbf{Power:} Due to its simple logic and the fact that it is only active during sparsely sampled epochs, the power consumption of the trace collector is negligible.
\end{itemize}

By writing the trace data to OS-managed system memory, our solution is not only efficient but also simplifies the data pipeline, as the trace becomes directly accessible by the GPU or our custom accelerator for inference.

\paragraph{Security and Privacy Considerations}
A trace collection mechanism that captures instruction and memory addresses is inherently sensitive. For the \textbf{Enterprise Forecaster}, trace confidentiality is paramount, and for \textbf{Ecosystem Partners}, data privacy is a primary concern. Our design addresses this by treating the trace buffer as a protected memory region, accessible only by a trusted driver and the inference engine. For heightened security, the trace data can be encrypted on-the-fly by the collector hardware before being written to system memory. This ensures trace opaqueness even to a compromised kernel. Furthermore, we emphasize that only metadata about the execution is captured; at no point is the raw data from memory or registers ever exposed in the trace.

\subsection{High-Speed Inference on Commodity GPUs}
A primary deployment target for \neuroscalar{} is the commodity GPU owing to its ubiquity 
and its mature parallel programming ecosystem. 
We detail our software-based inference engine, its performance characteristics, and the methodology for achieving transparent, low-overhead execution.

Our implementation is intentionally straightforward. The trace data for a sampled epoch, residing in OS-managed memory, is transferred to the GPU via standard user-space driver calls. The \neuroscalar{} model is then executed using a conventional deep learning inference framework. The only optimization we use is the quantization of model weights to \textbf{FP16}, which halves the memory footprint and bandwidth requirements without any measurable impact on prediction accuracy for this task.

The performance of our model on a GPU is dictated by the deliberate lightweight design. Table~\ref{tab:speed} shows measurements for inference speed of 1000 epochs on six different GPUs, measured in inferred instructions per second. As detailed earlier, \neuroscalar{} is an LSTM-based model with a hidden dimension of only 256. This means that the core computation is a series of relatively small matrix-matrix multiplications (GEMMs). For an epoch of 100,000 instructions, processed with a batch size of 100, the key operation is a matrix multiplication of roughly (100x1024) $\times$ (1024x1024) (applying a single matrix multiplication for the four f, g, i, o LSTM gates~\cite{hochreiter1997long}). As shown by prior work analyzing GPU kernel efficiency, such as ~\cite{10.1145/3620665.3640367}, GEMM operations with these dimensions are too small to saturate the memory bandwidth or fully utilize the TensorCores of a modern desktop GPU, sustaining only about 4\% of theoretical peak FLOPs.

However, this underutilization is a consequence of our goal of a lightweight model, and the resultant absolute performance is more than sufficient for our needs. On a commodity NVIDIA 4090 GPU, we achieve a simulation speed of \textbf{4 to 5 million instructions per second (MIPS)}. To translate this to user-felt overhead, we perform the following analysis:
\begin{itemize}
    \item Time to run GPU inference on one epoch (100,000 instructions) is: $100,000 \text{ inst} / 4,000,000 \text{ inst/sec}$ $ = 0.025s$.
    \item To maintain a system overhead of 0.1\%, the CPU must be allowed to execute for $0.025\text{s} / 0.001 = 25s$ in the time it takes to process one epoch.
    \item Assume a host CPU executes instructions at a rate of 3,000 MIPS (3 GHz, IPC=1).
    \item This means a sampling rate of one epoch for every $\sim$0.75 million epochs executed, which aligns with our system goals.
\end{itemize}

\begin{table}[tbp]
  \centering
  \caption{GPU inference speed (instructions/second).}
  \label{tab:speed}
  \setlength{\tabcolsep}{6pt}
  \begin{tabular}{lcc}
    \toprule
    GPU & Mean & Std.\ Dev. \\
    \midrule
    H100 & 4{,}846{,}724 & 522{,}613 \\
    A100 & 4{,}420{,}457 & 340{,}531 \\
    L40 & 5{,}309{,}333 & 294{,}540 \\
    L40S & 5{,}793{,}294 & 287{,}928 \\
    RTX 4090 & 5{,}916{,}285 & 362{,}703 \\
    RTX 5000 & 4{,}233{,}920 & 174{,}764 \\
    \bottomrule
  \end{tabular}
\end{table}




Finally, to ensure true transparency, the inference engine is designed to \textit{scavenge} GPU resources. The inference kernels are launched at a low OS priority, ensuring that any user-facing, latency-sensitive applications (e.g., gaming, UI rendering) are always given precedence. This prevents our background processing from introducing any noticeable stutter or lag. Once inference is complete, the resulting cycle predictions are either stored locally for the \textbf{Enterprise Forecaster's} internal use or aggregated and sent back to the chip designer. Whether the DL weights might leak some information of the microarchitecture is a valid concern. The DL weights themselves can be protected by encrypting and decrypting before use in a TEE environment and secure attestation modern GPUs provide. Finally, simply the availability of performance projection, might allow microbenchmark-based reverse engineering: to protect from this, the system stack can be designed such that the inference outputs are not visible to users.

\subsection{High-Speed Inference with an On-Chip Accelerator}
\begin{figure}
    \centering
    \includegraphics[width=0.9\linewidth]{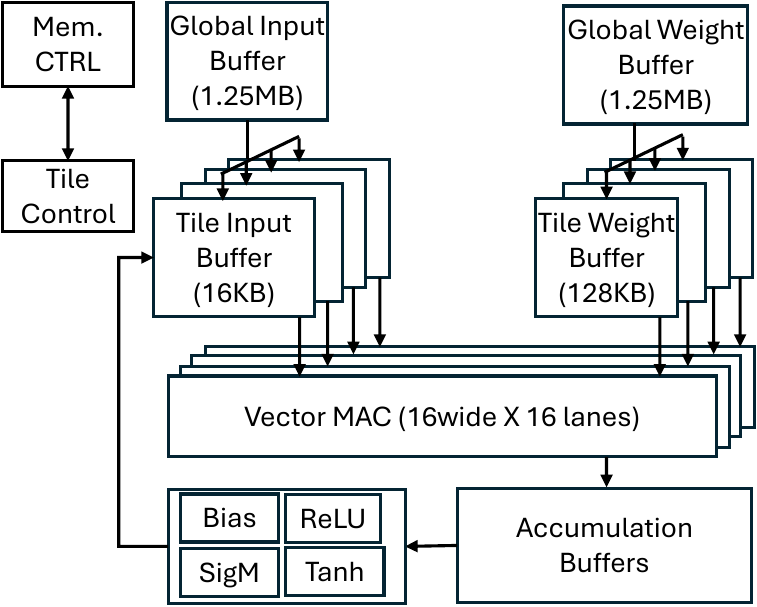}
    \caption{Neutrino Inference Accelerator.}
    \label{fig:accelerator-overview}
\end{figure}

While commodity GPUs offer a flexible and high-performance inference platform, they are not always available, particularly in power-constrained environments. To provide a dedicated, ultra-low-power alternative, we co-designed a hardware accelerator, dubbed Neutrino, specifically tailored to the \neuroscalar{} DL model. This section describes the accelerator's architecture, its dataflow, and the performance benefits derived from this holistic design approach.

\subsubsection{Accelerator Architecture and Dataflow}

The design of Neutrino is informed by the principles of successful deep learning hardware like NVDLA~\cite{nvdla}, MAGNET~\cite{venkatesan2019magnet}, and Eyeriss~\cite{chen2016eyeriss}. The core of the accelerator is a 256-wide \textbf{INT8 vector engine}, structured as 16x16 lanes as shown in Figure~\ref{fig:accelerator-overview}. This is supported by a hierarchical memory system designed to eliminate off-chip memory access during inference. An epoch's entire feature trace (quantized to 1MB) is first loaded into a \textbf{1.25MB on-chip SRAM buffer} from CPU memory. Weights are loaded once at boot-time into the global weight buffer of \textbf{1.25MB buffer}. From there, weights and activations are staged into smaller, local buffers: a \textbf{128KB weight buffer} and a \textbf{16KB input buffer} directly feeding the vector engine.

This on-chip memory hierarchy allows the accelerator's execution to be completely \textbf{statically scheduled}. Once an epoch's input features are loaded into the main SRAM buffer, the entire inference process proceeds without any stalls for DRAM access, enabling predictable, high-performance execution. The dataflow consists of staging weights from the larger weight buffer into the local accelerator buffers, performing vector operations, and accumulating partial sums.

\subsubsection{Co-Design for Maximum Utilization}

The novelty of this design lies in the tight \textbf{co-design of the DL model, the accelerator's microarchitecture, and the execution dataflow}. The model's hidden dimension of 256 was deliberately chosen to perfectly match the 256-wide vector engine. The dominant computation in our LSTM model is the vector-matrix multiplication required for the input, output, forget, and cell gates. For a single instruction trace, this becomes a (1, 256) $\times$ (256, 256) operation, which is executed on the 256-wide vector unit in 256 cycles. This precise matching allows the vector lanes to achieve nearly 100\% utilization during these matrix operations. The application of the bias, and non-linear sigmoid and tanh activation functions, which are orders of magnitude less computationally intensive, are temporally pipelined after the MATMUL. A single LSTM layer finishes in 8264 cycles, achieving 99\% MAC utilization.

\subsubsection{Performance and Efficiency}

To verify the design and obtain accurate performance and power figures, we developed a cycle-level simulator for the accelerator and a simple compiler to map the LSTM operations onto its instruction set. The simulator's performance was validated against a full RTL implementation of the accelerator, synthesized using a commercial 16nm P\&R flow. Results are in Table~\ref{tab:accel_metrics}, for a single-tile and an 8-tile design.

\begin{table}
    \centering
    \begin{tabular}{c|c|c}
     & 1 Tile & 8 Tiles\\ \hline
        Total Area & 2.04$mm^2$ & 3.15$mm^2$  \\
        Tile Area & 0.16 $mm^2$ & 1.26$mm^2$ \\
        Power & 28mW & 226 mW\\
        Speed & 0.02 million inst/sec & 0.157 million inst/sec\\
        Sampling rate & 1/152642
& 1/19080 \\
    \end{tabular}
    \caption{Accelerator Metrics. Total area includes area of the Global Input Buffer and Global Weight Buffer large SRAMs.}
    \label{tab:accel_metrics}
\end{table}

Our analysis shows that Neutrino can process traces at a rate of \textbf{0.02 million instructions per second (MIPS)} while consuming only \textbf{28 milliwatts} of power. This represents a nearly \textbf{85$\times$ improvement in energy efficiency} (inferences/sec/watt) compared to the NVIDIA 4090 GPU solution. This result demonstrates the main benefit of our co-design approach: by tailoring the hardware and software together, we can create a solution that is orders of magnitude more efficient than a general-purpose programmable one. This design can be trivially scaled-out in an 8-tile design, each tile processing a different batch, with no inter-tile communication, while sharing of the global input buffer SRAM and weights SRAM.


\subsection{Collecting Traces For Training}
\label{sec:cycle-gather}
The foundation of \neuroscalar{} is a model trained on data generated from a trusted, high-fidelity simulator. This section briefly outlines the process of collecting this training data.

The training process requires paired examples of key properties and output labels. We instrument a conventional cycle-level simulator---in our case, \textbf{Gem5}---to generate these pairs. For each instruction executed in a simulation, we collect the six key properties defined in Section 4.1 (PC, memory address, etc.). It is critical that the properties extracted from the simulator are \textbf{semantically identical} to those captured by the hardware tracer to prevent any train-serve skew.

The corresponding ground-truth label for each instruction is its \textbf{retirement latency}: the number of cycles elapsed since the previous instruction retired. This cycle count is a direct output of the simulator's detailed pipeline model. To create a robust and generalizable foundation model, we generate our training corpus by running a diverse set of workloads, including 16 applications from the SPEC CPU 2017 benchmark suite, through our instrumented simulator.

While our core methodology relies on microarchitecture-independent features to enable in-the-wild deployment on existing hardware, the framework is extensible. For scenarios where a designer wishes to model features of a new ISA or a specific hardware mechanism, those microarchitecture-dependent features could be added to the training input to potentially increase model accuracy further.

\subsection{Deployment and Model Management}

A successful in-the-wild simulation system requires more than just hardware and models; it needs a robust infrastructure for software control, model distribution, and data reporting. This section outlines the complete deployment and management framework for \neuroscalar{}.

\subsubsection{Software Interface and Control}

End-users interact with the \neuroscalar{} system through a dedicated driver and a user-space API. This interface provides the necessary control to manage the tracing and inference process, critical for the  \textbf{Enterprise Forecaster} user. The driver is responsible for configuring the hardware trace collector, managing the secure memory buffer, and scheduling the inference tasks on the GPU or accelerator, as described in Section 4.2. It ensures that tracing is properly synchronized with the OS scheduler to handle context switches and maintain process isolation. If application slowdown is permitted, sampling frequency can be tuned as well.

%

\subsubsection{Model Distribution and Updating}

The chip designer makes \neuroscalar{} models, each representing a different microarchitectural design, available through a secure \textbf{model repository}. When a user requests a specific model, it is downloaded, cryptographically verified, and stored locally. This repository-based approach allows for seamless updates and version control. If a designer retrains a model with more data to improve its accuracy, end-users can be prompted to update to the latest version, ensuring they are always working with the most current and accurate performance predictions.

\subsubsection{Telemetry and Reporting Framework}

For the \textbf{Ecosystem Partner} use case, a secure and privacy-preserving telemetry pipeline is essential for returning performance data to the chip designer. The process is designed to be fully transparent and opt-in. When inference for a sampled epoch is complete, the result (a predicted IPC or cycle count) is aggregated locally. Periodically, this aggregated, anonymized data---stripped of any personally identifiable information---is sent back to a central collection server managed by the chip designer. The data format includes the model version, the application signature (SHA256 hash of PCs), and the performance prediction. This allows designers to build a large-scale, real-world dataset of how their architectural ideas perform across thousands of applications and usage scenarios. 

This ecosystem can extend existing logging frameworks like those developed by hyperscalars such as Google's OpenTelemetry~\cite{opentelemetry} and Meta's Dynolog~\cite{dynolog}, or commercial solutions from providers like Datadog~\cite{datadog} and Splunk~\cite{splunk}. While these platforms are typically geared towards software and infrastructure monitoring, our hardware-derived performance predictions could be integrated as a novel telemetry source, providing chip designers with direct, in-the-wild feedback through established and trusted data pipelines.
NVIDIA's GeForce Telemetry~\cite{geforce-experience}, Intel’s Platform Monitoring Technology (PMT) and Telemetry Interface Specification (TIS) allow telemetry data to be collected already with a full-fledged system infrastrcture in place - performance projection is simply another source of telemetry data.

\section{Experimental Methodology}

\newcommand{\base}[0]{8w}
\newcommand{\slow}[0]{6w+ls}
\newcommand{\ew}[0]{rob}
\newcommand{\mem}[0]{lsq}
\newcommand{\fourw}[0]{4w+mem}

\paragraph{Model construction}
To construct the DL model we used the GEM5 simulator which was annotated to produce the feature trace and the ground truth retirement cycle count for each instruction. We did this on 16 SPEC2017 benchmarks. We fast-forwarded by 2 billion instructions and created a dataset comprising of 100 million instructions for each application. We also studied four additional configurations besides our baseline \base configuration. The details of the configurations are shown in Table~\ref{tab:proc_configs}. For inference testing, 10 million instructions from this dataset are set aside and never used for training. The model training and experiments were done on a mix of A100, L40, L40S, and A10. 
The final unified model trained on the \base configuration, across all 16 benchmarks took a total of 8 hours to train on an A100.

\begin{table}[t]
  \centering
  \caption{Microarchitectural parameters for the five processor configurations evaluated. The baseline is an 8-wide out-of-order processor. L2 cache is 8MB across the board. Variations explore different trade-offs in memory and core resources.}
  \renewcommand{\arraystretch}{0.75}
  \label{tab:proc_configs}
  \begin{tabular}{lccccc}
    \toprule
    \multirow{2}{*}{\textbf{Config}}
     & \textbf{Base} & \textbf{6-wide+} & \textbf{Large} & \textbf{Large} & \textbf{More} \\
     & \textbf{8-wide}&  \textbf{LS Unit} & \textbf{ROB} & \textbf{LSQ} & \textbf{Memory} \\
     Abbr. & \textit{(\base)} & \textit{(\slow)} & \textit{(\ew)} & \textit{(\mem)} & \textit{(\fourw)} \\
    \midrule
    Width & 8 & 6 & 8 & 8 & 4 \\
    LS Units & 1 & 2 & 1 & 2 & 2 \\
    LSQ Entries & 32 & 32 & 32 & 64 & 32 \\
    Num Regs & 256 & 256 & 512 & 256 & 256 \\
    ROB Size & 192 & 192 & 384 & 192 & 192 \\
    \midrule
    L1D\$ Size & 64KB & 64KB & 64KB & 64KB & 128KB \\
    L1I\$ Size & 64KB & 64KB & 64KB & 64KB & 64KB \\
    \bottomrule
  \end{tabular}
\end{table}

\paragraph{Inference} To study inference we run the model on a family of commodity GPUs and measure speed (inferences/second). The platforms we considered are A100, H100, A6000, 4090, and an RTX5000. Our accelerator was implemented in RTL and synthesized with a full  P\&R flow at 16nm. Recall that our accelerator implements a complete static schedule, and its execution is deterministic in the number of cycles for an epoch (with any potential delays only arising from the time to transfer from the processor's memory into the accelerator's Global Input Buffer SRAM). The accelerator's main results are summarized in Table~\ref{tab:accel_metrics}.

\paragraph{Downstream tasks} We examine two downstream tasks for a chip designer as case studies. Starting from an 8w processor, the chip designer seeks to understand how performance changes for 4 other configurations. In particular, they seek to understand at the instruction level. We report two studies where the 5 processors are ranked for each retiring instruction for each benchmark. Second, we present pairwise comparison for each benchmark for each instruction. This allows extremely low-level analysis on in-the-wild workloads. Both of these push the model's accuracy and fidelity to the limit - the changes between the processor configurations is intentionally set to subtle features like LSQ size, \# ports etc.

\section{Results}

First, we show that individual benchmark models work well and can transfer to other models. Second, a model is examined that uses information from every benchmark. Finally, we test the accuracy in downstream tasks like pairwise ordering and best-configuration. 

\subsection{Benchmark-wise Error and Accuracy Heatmaps}
We visualize cross-benchmark behavior with two heatmaps. In both plots, \emph{rows} denote the \textbf{source} benchmark used to train/fit the predictor and \emph{columns} denote the \textbf{target} benchmark on which we evaluate. Each cell is annotated with the corresponding score for that (source, target) pair.

\begin{figure}[t]
  \centering
  \includegraphics[width=0.9\columnwidth]{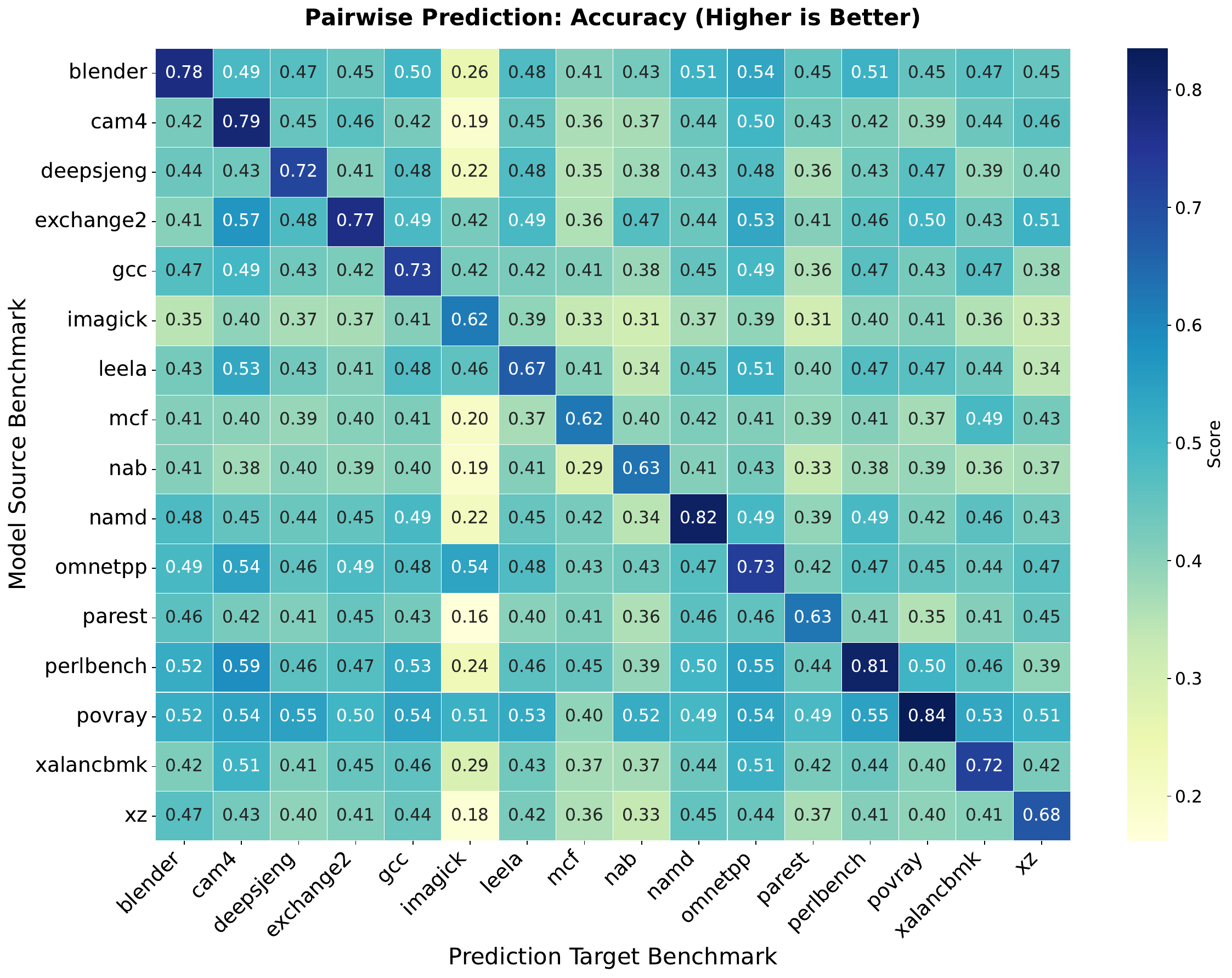}
  \caption{Pairwise prediction \textbf{Acc(round)}}
  \label{fig:acc-heatmap}
\end{figure}

\paragraph{Prediction accuracy.}
Figure~\ref{fig:acc-heatmap} shows \emph{Acc(round)} results. In-domain accuracies exceed $70\%$ (often $80\%+$), while cross-dataset evaluations remain around $50\%$, demonstrating reasonable generalization. Overall, $\sim80\%$ of instructions are predicted exactly and over $95\%$ within $\pm1$ cycle, with a detailed error figure deferred to Appendix~\ref{sec:appendix}.

\subsection{Foundation Model on the Concatenated Corpus}
We train a single backbone on the union of all benchmark splits (\textbf{TraceFusion-13}) and evaluate it on held-out data. The model achieves consistently strong accuracy across diverse workloads, generalizing well to both short-latency and long-tail benchmarks. Table~\ref{tab:uni-foundation-acc} summarizes representative results, with full details in Appendix~\ref{sec:appendix}.

\begin{table}[t]
  \centering
  \small
  \caption{Accuracy-centric evaluation of the foundation model (\textbf{TraceFusion-13})}
  \label{tab:uni-foundation-acc}
  \setlength{\tabcolsep}{6pt}
  \renewcommand{\arraystretch}{0.9}
  \begin{tabular}{lccc}
    \toprule
    Benchmark & $\pm1\uparrow$ & $\pm2\uparrow$ & Acc(round)\,(ref)$\uparrow$ \\
    \midrule
    \texttt{xz}         & 0.8388 & 0.9159 & 0.6782 \\
    \texttt{gcc}        & 0.8607 & 0.9378 & 0.7174 \\
    \texttt{xalancbmk}  & 0.8739 & 0.9415 & 0.7231 \\
    \texttt{nab}        & 0.7802 & 0.8947 & 0.6232 \\
    \texttt{deepsjeng}  & 0.8579 & 0.9430 & 0.7102 \\
    \texttt{parest}     & 0.8002 & 0.8999 & 0.6246 \\
    \texttt{namd}       & 0.9391 & 0.9725 & 0.8116 \\
    \texttt{povray}     & 0.9545 & 0.9821 & 0.8318 \\
    \bottomrule
  \end{tabular}
\end{table}

\noindent\emph{Summary: Neuroscalar is highly accurate and at fine-granularity of individual instructions.}

\subsection{Downstream Tasks: Processor-Config Comparisons}

\paragraph{Pairwise ordering accuracy}
For each configuration pair $(i \le j)$, we evaluate whether the predicted ordering preserves the ground-truth ordering. 
Table~\ref{tab:pairwise-full} reports aggregated statistics across all benchmarks. 
In addition to the overall correct/total counts dubbed match rate, we also include the fraction of cases where the ground truth indicates $i$ is strictly better than $j$ (\texttt{GT-better}) and the fraction of samples with non-zero ground-truth cycles (\texttt{Non-zero}). 
These richer statistics provide a more complete picture of pairwise ordering fidelity, showing that pairwise orderings are preserved at roughly $90\%$ rates depending on the configuration pair. The 3rd column shows the ``irregular'' nature of OOO processors: comparing \fourw $\le$ \base, for 15\% of instructions, the \fourw \space processors retires earlier than the \base. Neuroscalar is able to learn and capture this behavior as well.

\begin{table}[t]
  \centering
  \caption{Pairwise ordering results aggregated across benchmarks. Columns: match rate, proportion of ground-truth cases where the left config is strictly better (\texttt{GT-better}), and proportion of non-zero ground-truth samples (\texttt{Non-zero}).}
  \label{tab:pairwise-full}
  \setlength{\tabcolsep}{4pt}
  \renewcommand{\arraystretch}{0.95}
  \begin{tabular}{rclccc}
    \toprule
    \multicolumn{3}{c}{Pair} & Match rate & GT-better & Non-zero \\
    \midrule
    \fourw&$\le$&\base        & $91.11\%$ & $15.0\%$ & $39.2\%$ \\
    \fourw&$\le$&\ew & $91.08\%$ & $15.0\%$ & $39.3\%$ \\
    \fourw&$\le$&\mem       & $91.06\%$ & $15.0\%$ & $39.4\%$ \\
    \fourw&$\le$&\slow      & $91.04\%$ & $15.1\%$ & $39.5\%$ \\
    \base&$\le$&\ew & $91.06\%$ & $15.0\%$ & $39.4\%$ \\
    \base&$\le$&\mem       & $91.08\%$ & $15.0\%$ & $39.3\%$ \\
    \base&$\le$&\slow      & $91.11\%$ & $15.0\%$ & $39.3\%$ \\
    \ew&$\le$&\mem & $91.14\%$ & $15.0\%$ & $39.2\%$ \\
    \ew&$\le$&\slow & $91.17\%$ & $15.0\%$ & $39.2\%$ \\
    \mem&$\le$&\slow     & $91.20\%$ & $15.0\%$ & $39.1\%$ \\
    \bottomrule
  \end{tabular}
\end{table}

\paragraph{Five-config ranking: full-rank match, best-config match, and Kendall $\tau$}
For each instruction, we rank five configs by rounded cycles and report: (i) full permutation match, (ii) whether the best (lowest-cycle) config is matched, and (iii) Kendall $\tau$ over all $\binom{5}{2}$ pairs with ties handled by group ranks. Representative end-of-run summaries is show in Table~\ref{tab:five-config-ranking}, where Kendall~$\tau$ typically falls between $0.8$ and $0.95$, and best-config match consistently exceeds $94\%$.

\noindent\emph{Summary: Neuroscalar can extract detailed per-instruction differences between different microarchitectures accurately.}

\begin{table}[t]
  \centering
  \small
  \setlength{\tabcolsep}{4pt}
  \renewcommand{\arraystretch}{0.95}
  \caption{Five-config ranking per benchmark. Metrics over held-out instructions: Kendall $\tau$ (higher is better), full permutation match, and best-config match.}
  \label{tab:five-config-ranking}
  \begin{tabular}{lccc}
    \toprule
    \textbf{Benchmark} & \textbf{Kendall $\tau$} & \textbf{Full Match(\%)} & \textbf{Best Match(\%)} \\
    \midrule
    xz          & 0.8657 & 56.93 & 96.09 \\
    gcc         & 0.8449 & 52.31 & 96.24 \\
    xalancbmk   & 0.8036 & 48.35 & 96.46 \\
    nab         & 0.8188 & 50.88 & 94.67 \\
    deepsjeng   & 0.8360 & 53.87 & 96.21 \\
    parest      & 0.7741 & 43.92 & 94.24 \\
    namd        & 0.9150 & 64.45 & 98.19 \\
    povray      & 0.9355 & 64.20 & 98.61 \\
    \bottomrule
  \end{tabular}
\end{table}




\section{Related Work}
Prior work in applying machine learning to performance prediction has critical limitations for in-the-wild deployment. Early efforts like Ithemal~\cite{mendis2019ithemal} focused on predicting the latency of static basic blocks, but by ignoring dynamic execution phenomena like memory access and branch prediction outcomes, they cannot model realistic, full-program behavior. More recent full-program simulators fall into two categories, neither of which is suitable for our goals. SimNet~\cite{10.1145/3530891} achieves accuracy by using microarchitectural ``context'' features---such as which level of the cache hierarchy served a request. This approach is fundamentally incompatible with in-the-wild prediction, as these features are the \textit{output} of a specific microarchitecture and cannot be collected from production hardware for a hypothetical design. Conversely, TAO~\cite{10.1145/3656012} builds a large, monolithic model to explore a predefined space of \textit{basic} parameters (e.g., ROB size). This makes it too computationally heavyweight for lightweight, sampling-based deployment and, more importantly, too inflexible to evaluate the novel, complex architectural ideas (like a new prefetcher) that designers need to test. \textit{In summary, previous methods are either too limited in scope, depend on features unavailable on production hardware, or are too monolithic to evaluate novel design trade-offs.} Our work provides a lightweight, extensible framework specifically designed to fill this gap. 

\textbf{ML-based performance models.} ML-based performance models apply/use ideas from linear regression and interpolation, to extrapolate performance of new microarchitecture from sampling the space. They lack the ability in general for fine-grained microarchitecture policies and typically fail to accurately model the run-time complex dynamic interaction between the program and hardware~\cite{li2009machine,zheng2016accurate,ipek2006efficiently,joseph2006construction,joseph2006predictive,lee2006accurate,lee2007illustrative}.
ML models have also been developed for application performance prediction across ISAs and machine architecture~\cite{zheng2016accurate, zheng2015learning,ardalani2015cross,baldini2014predicting,o2017gpu}. Concorde is a state-of-art hybrid ML/analytical model~\cite{10.1145/3695053.3731037} - for in-the-wild simulation it exposes too much processor information.



\section{Conclusion}
This paper studies a fundamental bottleneck in computer architecture research: the inability to evaluate novel microarchitectural ideas with both high fidelity and high speed on real-world, "in-the-wild" workloads. Traditional cycle-accurate simulation is too slow, and existing ML-based approaches are either incompatible with in-field deployment or too high-level to capture the detailed design trade-offs that matter to practitioners. Our solution is a deep learning framework built on the key insight of using \textbf{microarchitecture-independent features} to train an ultra-lightweight DL model. This approach decouples performance prediction from the underlying hardware, allowing a model to run on existing silicon, extract features of a workload, and accurately forecast it's performance on a hypothetical design. We presented a complete system, including a lightweight hardware tracer, a sampling methodology, commodity GPU-based inference and an ultra-low-power accelerator, that performs this analysis with negligible overhead. Our evaluation shows that this framework can differentiate between competing designs with \textbf{95\% accuracy} in A/B testing scenarios, providing architects with a powerful new tool for data-driven design. This work paves the way for a future where hardware design cycles are dramatically accelerated through large-scale, continuous feedback from live user workloads.

\clearpage
\bibliographystyle{plain}
\bibliography{references,references_penrose}

\appendix
\section{Appendix}

\label{sec:appendix}
We provide results that expand on those presented in the main paper. These are provided for completeness.

\begin{table*}[t]
  \centering
  \caption{Detailed parameter shapes and counts of the proposed model.}
  \label{tab:model-params}

  \begin{tabular}{lcc}
    \hline
    \textbf{Parameter} & \textbf{Shape} & \textbf{\#Params} \\
    \hline
    input\_proj.weight          & [256, 13]   & 3,328 \\
    input\_proj.bias            & [256]       & 256 \\
    rnn.weight\_ih\_l0          & [1024, 256] & 262,144 \\
    rnn.weight\_hh\_l0          & [1024, 256] & 262,144 \\
    rnn.bias\_ih\_l0            & [1024]      & 1,024 \\
    rnn.bias\_hh\_l0            & [1024]      & 1,024 \\
    rnn.weight\_ih\_l1          & [1024, 256] & 262,144 \\
    rnn.weight\_hh\_l1          & [1024, 256] & 262,144 \\
    rnn.bias\_ih\_l1            & [1024]      & 1,024 \\
    rnn.bias\_hh\_l1            & [1024]      & 1,024 \\
    cls\_fc1.weight             & [64, 256]   & 16,384 \\
    cls\_fc1.bias               & [64]        & 64 \\
    cls\_fc2.weight             & [2, 64]     & 128 \\
    cls\_fc2.bias               & [2]         & 2 \\
    output\_layer\_short.weight & [1, 256]    & 256 \\
    output\_layer\_short.bias   & [1]         & 1 \\
    output\_layer\_long.weight  & [1, 256]    & 256 \\
    output\_layer\_long.bias    & [1]         & 1 \\
    \hline
    \textbf{Total}              & --          & 1,073,348 \\
    \hline
  \end{tabular}
\end{table*}


\begin{table*}[t]
  \centering
  \caption{Foundation model (\textbf{TraceFusion-13}) evaluated per benchmark (held-out splits). Error-centric metrics are emphasized; \textit{Acc(round)} is reported for reference. All numbers are fractions except MAE/RMSE (cycles).}
  \label{tab:uni-foundation}
  \begin{tabular}{lccccccc}
    \toprule
    Benchmark & rel$\le$5\%\,$\uparrow$ & RAE$\downarrow$ & MAE$\downarrow$ & RMSE$\downarrow$ & $\pm1\uparrow$ & $\pm2\uparrow$ & Acc(round)\,(ref)$\uparrow$ \\
    \midrule
    \texttt{blender}    & 0.7791 & 0.1862 & 0.5795 & 4.0245 & 0.9158 & 0.9646 & 0.7774 \\
    \texttt{cam4}       & 0.7941 & 0.1367 & 0.3762 & 4.9745 & 0.9490 & 0.9803 & 0.7913 \\
    \texttt{deepsjeng}  & 0.7104 & 0.2580 & 0.9147 & 6.3028 & 0.8579 & 0.9430 & 0.7102 \\
    \texttt{exchange2}  & 0.7614 & 0.2016 & 0.6695 & 3.9367 & 0.8960 & 0.9579 & 0.7599 \\
    \texttt{gcc}        & 0.7189 & 0.2650 & 0.9408 & 7.5068 & 0.8607 & 0.9378 & 0.7174 \\
    \texttt{imagick}    & 0.6098 & 0.3425 & 0.8163 & 2.2620 & 0.7827 & 0.9043 & 0.6098 \\
    \texttt{leela}      & 0.6714 & 0.3358 & 1.2239 & 6.1895 & 0.8163 & 0.9160 & 0.6704 \\
    \texttt{mcf}        & 0.6174 & 0.3795 & 2.4035 & 23.5354 & 0.7768 & 0.8997 & 0.6171 \\
    \texttt{nab}        & 0.6240 & 0.3844 & 1.4612 & 6.7714 & 0.7802 & 0.8947 & 0.6232 \\
    \texttt{namd}       & 0.8156 & 0.1415 & 0.4563 & 4.2321 & 0.9391 & 0.9725 & 0.8116 \\
    \texttt{omnetpp}    & 0.7263 & 0.2539 & 0.9170 & 5.9410 & 0.8637 & 0.9377 & 0.7244 \\
    \texttt{parest}     & 0.6252 & 0.5029 & 3.5813 & 34.1094 & 0.8002 & 0.8999 & 0.6246 \\
    \texttt{perlbench}  & 0.8064 & 0.1347 & 0.3520 & 4.8300 & 0.9570 & 0.9836 & 0.8054 \\
    \texttt{povray}     & 0.8333 & 0.1140 & 0.3143 & 1.7281 & 0.9545 & 0.9821 & 0.8318 \\
    \texttt{xalancbmk}  & 0.7263 & 0.2454 & 0.9135 & 8.8675 & 0.8739 & 0.9415 & 0.7231 \\
    \texttt{xz}         & 0.6790 & 0.3386 & 1.2454 & 6.5660 & 0.8388 & 0.9159 & 0.6782 \\
    \bottomrule
  \end{tabular}
\end{table*}

\begin{table*}[t]
  \centering
  \caption{Pairwise ordering per benchmark (Part I). Each cell shows \emph{Match rate / GT-better / Non-zero}.}
  \label{tab:pairwise-per-benchmark-part1}
  \setlength{\tabcolsep}{4pt}
  \renewcommand{\arraystretch}{0.92}
  \resizebox{\textwidth}{!}{
  \begin{tabular}{lccccc}
    \toprule
    Benchmark & \fourw $\le$ \base & \fourw $\le$ \ew & \fourw $\le$ \mem & \fourw $\le$ \slow & \base $\le$ \ew \\
    \midrule
    blender   & 92.75 / 13.80 / 39.21 & 92.73 / 13.81 / 39.30 & 92.72 / 13.81 / 39.39 & 92.72 / 13.83 / 39.48 & 92.73 / 13.76 / 39.41 \\
    cam4      & 96.46 / 16.28 / 35.99 & 96.45 / 16.29 / 36.05 & 96.47 / 16.29 / 36.12 & 96.49 / 16.30 / 36.18 & 96.48 / 16.25 / 36.14 \\
    deepsjeng & 90.49 / 14.75 / 39.38 & 90.45 / 14.77 / 39.47 & 90.41 / 14.76 / 39.57 & 90.38 / 14.79 / 39.66 & 90.41 / 14.74 / 39.60 \\
    exchange2 & 92.74 / 14.70 / 39.19 & 92.75 / 14.71 / 39.28 & 92.76 / 14.69 / 39.37 & 92.77 / 14.70 / 39.46 & 92.80 / 14.62 / 39.39 \\
    gcc       & 90.00 / 16.00 / 38.79 & 89.99 / 16.03 / 38.87 & 89.97 / 16.02 / 38.95 & 89.97 / 16.03 / 39.02 & 89.98 / 15.95 / 38.98 \\
    imagick   & 81.62 / 15.33 / 34.47 & 81.40 / 15.31 / 34.61 & 81.18 / 15.29 / 34.75 & 81.01 / 15.30 / 34.88 & 81.14 / 15.27 / 34.79 \\
    leela     & 89.80 / 15.12 / 38.53 & 89.77 / 15.12 / 38.64 & 89.74 / 15.11 / 38.74 & 89.72 / 15.13 / 38.83 & 89.75 / 15.07 / 38.77 \\
    mcf       & 89.36 / 15.99 / 43.49 & 89.29 / 16.05 / 43.61 & 89.22 / 16.08 / 43.73 & 89.16 / 16.14 / 43.85 & 89.17 / 16.04 / 43.83 \\
    nab       & 89.00 / 14.85 / 34.70 & 88.98 / 14.89 / 34.81 & 88.96 / 14.87 / 34.91 & 88.94 / 14.89 / 35.02 & 89.01 / 14.82 / 35.05 \\
    namd      & 94.37 / 15.04 / 40.05 & 94.37 / 15.06 / 40.15 & 94.37 / 15.06 / 40.24 & 94.38 / 15.09 / 40.32 & 94.38 / 15.02 / 40.26 \\
    omnetpp   & 89.45 / 15.83 / 37.66 & 89.45 / 15.86 / 37.74 & 89.46 / 15.86 / 37.83 & 89.47 / 15.89 / 37.91 & 89.48 / 15.82 / 37.87 \\
    parest    & 88.52 / 16.38 / 43.81 & 88.50 / 16.40 / 43.86 & 88.48 / 16.41 / 43.92 & 88.47 / 16.45 / 43.97 & 88.48 / 16.35 / 43.92 \\
    perlbench & 96.29 / 15.16 / 37.91 & 96.28 / 15.16 / 38.02 & 96.28 / 15.13 / 38.13 & 96.28 / 15.15 / 38.24 & 96.27 / 15.10 / 38.17 \\
    povray    & 95.09 / 14.78 / 33.47 & 95.09 / 14.78 / 33.56 & 95.08 / 14.77 / 33.64 & 95.08 / 14.79 / 33.72 & 95.08 / 14.75 / 33.66 \\
    xalancbmk & 89.72 / 15.50 / 34.52 & 89.72 / 15.50 / 34.64 & 89.72 / 15.47 / 34.76 & 89.73 / 15.47 / 34.88 & 89.76 / 15.39 / 34.88 \\
    xz        & 92.00 / 14.58 / 41.66 & 91.95 / 14.59 / 41.75 & 91.92 / 14.59 / 41.83 & 91.89 / 14.65 / 41.91 & 91.91 / 14.56 / 41.85 \\
    \bottomrule
  \end{tabular}}
\end{table*}

\begin{table*}[t]
  \centering
  \caption{Pairwise ordering per benchmark (Part II). Each cell shows \emph{Match rate / GT-better / Non-zero}.}
  \label{tab:pairwise-per-benchmark-part2}
  \setlength{\tabcolsep}{4pt}
  \renewcommand{\arraystretch}{0.92}
  \resizebox{\textwidth}{!}{
  \begin{tabular}{lccccc}
    \toprule
    Benchmark & \base $\le$ \mem & \base $\le$ \slow & \ew $\le$ \mem & \ew $\le$ \slow & \mem $\le$ \slow \\
    \midrule
    blender   & 92.76 / 13.72 / 39.34 & 92.79 / 13.75 / 39.28 & 92.82 / 13.72 / 39.21 & 92.86 / 13.75 / 39.15 & 92.89 / 13.78 / 39.08 \\
    cam4      & 96.46 / 16.22 / 36.11 & 96.47 / 16.25 / 36.07 & 96.46 / 16.22 / 36.03 & 96.47 / 16.25 / 36.00 & 96.50 / 16.28 / 35.92 \\
    deepsjeng & 90.44 / 14.71 / 39.54 & 90.46 / 14.74 / 39.48 & 90.50 / 14.72 / 39.42 & 90.52 / 14.75 / 39.36 & 90.55 / 14.77 / 39.29 \\
    exchange2 & 92.82 / 14.60 / 39.32 & 92.85 / 14.62 / 39.25 & 92.88 / 14.60 / 39.18 & 92.92 / 14.63 / 39.11 & 92.94 / 14.63 / 39.01 \\
    gcc       & 89.99 / 15.92 / 38.93 & 90.00 / 15.95 / 38.89 & 90.02 / 15.93 / 38.84 & 90.04 / 15.95 / 38.80 & 90.07 / 15.97 / 38.71 \\
    imagick   & 81.25 / 15.24 / 34.71 & 81.34 / 15.26 / 34.62 & 81.47 / 15.26 / 34.53 & 81.61 / 15.29 / 34.43 & 81.73 / 15.32 / 34.35 \\
    leela     & 89.76 / 15.04 / 38.70 & 89.78 / 15.08 / 38.64 & 89.81 / 15.06 / 38.57 & 89.83 / 15.09 / 38.51 & 89.87 / 15.11 / 38.42 \\
    mcf       & 89.17 / 16.00 / 43.82 & 89.17 / 16.02 / 43.81 & 89.18 / 15.98 / 43.80 & 89.18 / 15.99 / 43.78 & 89.18 / 16.01 / 43.75 \\
    nab       & 89.04 / 14.80 / 35.09 & 89.07 / 14.82 / 35.12 & 89.11 / 14.79 / 35.14 & 89.14 / 14.82 / 35.17 & 89.21 / 14.83 / 35.12 \\
    namd      & 94.40 / 14.98 / 40.21 & 94.42 / 15.00 / 40.15 & 94.44 / 14.97 / 40.10 & 94.47 / 15.00 / 40.04 & 94.49 / 15.01 / 39.97 \\
    omnetpp   & 89.51 / 15.79 / 37.82 & 89.55 / 15.82 / 37.78 & 89.58 / 15.79 / 37.73 & 89.62 / 15.82 / 37.69 & 89.66 / 15.84 / 37.62 \\
    parest    & 88.47 / 16.30 / 43.87 & 88.47 / 16.32 / 43.82 & 88.48 / 16.28 / 43.77 & 88.50 / 16.30 / 43.72 & 88.52 / 16.33 / 43.65 \\
    perlbench & 96.27 / 15.09 / 38.10 & 96.28 / 15.12 / 38.03 & 96.30 / 15.12 / 37.96 & 96.32 / 15.16 / 37.89 & 96.34 / 15.16 / 37.79 \\
    povray    & 95.09 / 14.73 / 33.60 & 95.11 / 14.77 / 33.55 & 95.12 / 14.74 / 33.50 & 95.14 / 14.77 / 33.45 & 95.15 / 14.81 / 33.37 \\
    xalancbmk & 89.79 / 15.36 / 34.88 & 89.83 / 15.38 / 34.88 & 89.87 / 15.35 / 34.88 & 89.90 / 15.37 / 34.88 & 89.95 / 15.37 / 34.85 \\
    xz        & 91.96 / 14.48 / 41.78 & 91.97 / 14.53 / 41.72 & 92.03 / 14.45 / 41.65 & 92.05 / 14.51 / 41.59 & 92.06 / 14.56 / 41.53 \\
    \bottomrule
  \end{tabular}}
\end{table*}

\begin{figure*}[t]
  \centering
  \includegraphics[width=\textwidth]{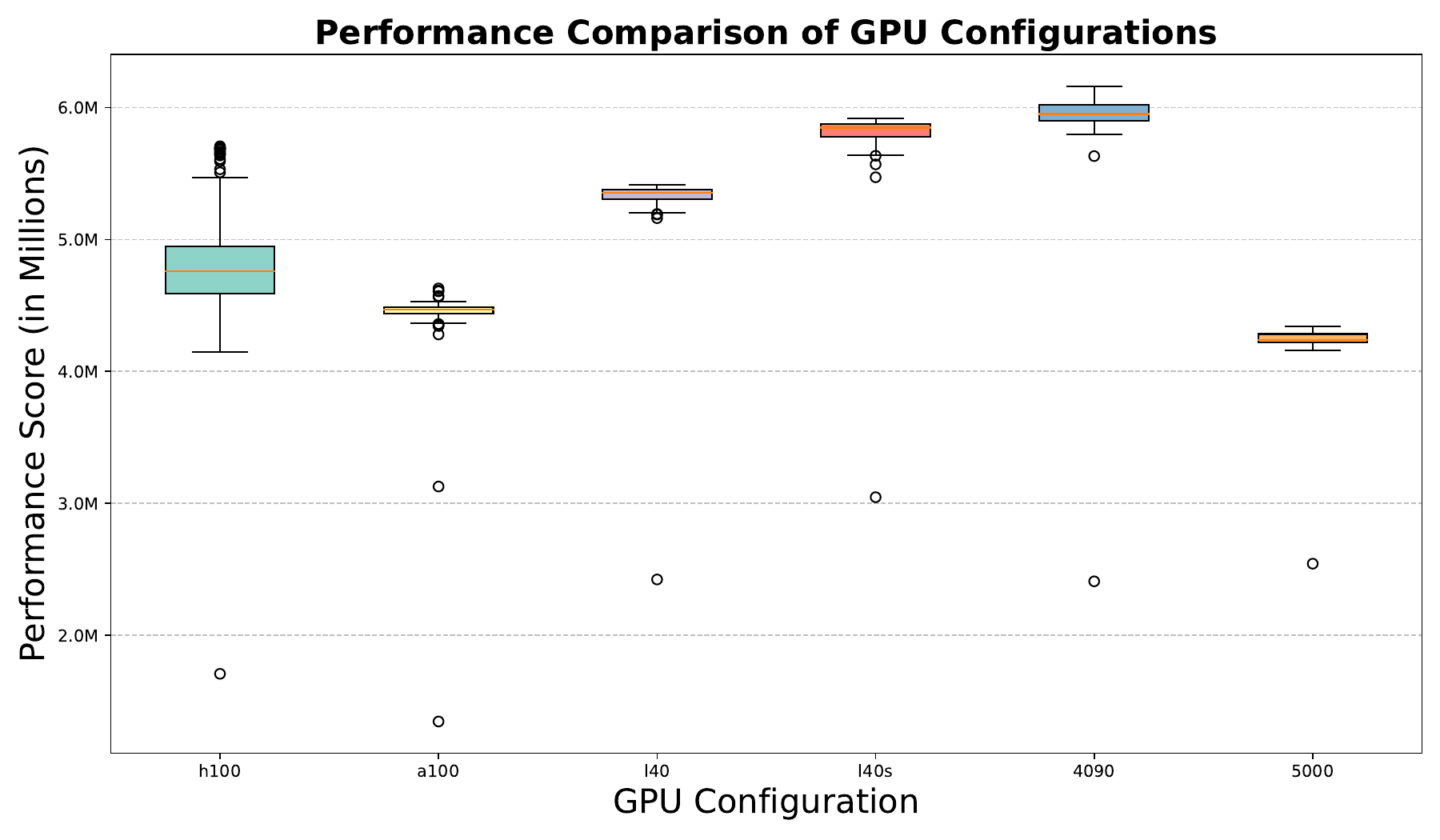}
  \caption{Distribution of GPU throughput across repeated runs. Boxplot visualization complements Table~\ref{tab:speed}, showing variability and stability of inference speeds.}
  \label{fig:gpu-box}
\end{figure*}

\begin{figure*}[t]
  \centering
  \includegraphics[width=\textwidth]{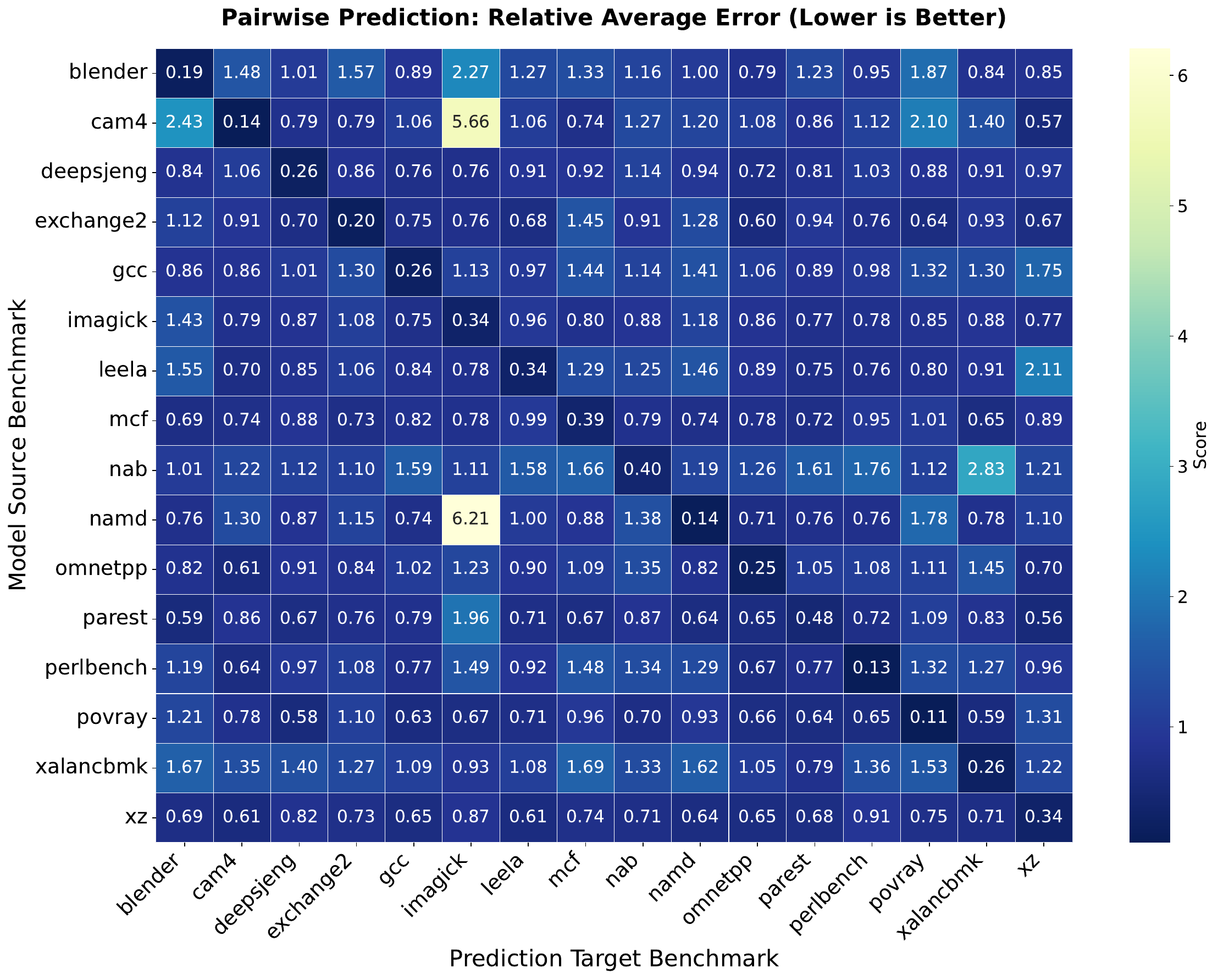}
  \caption{Pairwise prediction \textbf{Relative Average Error}.}
  \label{fig:err-heatmap}
\end{figure*}

\end{document}